\documentclass[a4paper, 11pt]{article}
\usepackage{comment} 
\usepackage{lipsum} 
\usepackage{fullpage} 
\usepackage{amsthm} 
\usepackage{csquotes} 
\usepackage{xcolor} 
\usepackage{graphicx} 
\usepackage[inline]{enumitem} 

\newtheorem*{personal}{\colorbox{blue!30}{Personal Data}}
\newtheorem*{consent}{\colorbox{blue!30}{Consent}}
\newtheorem*{profiling}{\colorbox{blue!30}{Profiling}}

\begin{document}
\noindent
\large\textbf{EU General Data Protection Regulation (GDPR)} \hfill \textbf{Sanchit Alekh} \\
\normalsize A Gentle Introduction \hfill \today \\

\section*{What is the GDPR and what does it aim to achieve?}
The \textit{GDPR}, or the \textit{Datenschutz Grundverordnung (DSGVO)} in German, is an EU Law which addresses the subject of safeguarding privacy of personal data of the citizens of the EU and EEA. It also specifies how data the collected data might be transported out of the EU/EEA. It is the first genuine effort to unify the plethora of disparate privacy regulations put forward by different regulatory bodies.

The GDPR aims to not only give more control over their personal data to the citizens, but also make conformance for businesses easier by defining unified guidelines. It also presses businesses, especially those dealing with sensitive personal data, to build their information systems in a way that confirms with \textit{Privacy by Design}. These regulations aim to ensure a more transparent handling and processing of personal data, and create an environment of trust and awareness on both sides, i.e., the data owner as well as the controllers/processors.

GDPR mandates that the highest privacy settings be used by default, and \enquote*{explicit and verifiable} consent be taken from the subject, and at the same time, also bestows the \textit{Right to Access} and the \textit{Right to Erasure}, which enables more transparency and accountability while handling personal data. The penalties for non-conformance are major enough that all kinds of companies, from small start-ups to large corporations, must take efforts and steps to follow these regulations.

\section*{Selected Definitions}
The GDPR provides unambiguous definitions for some important terms related to privacy. This is extremely important, because privacy, being something which is inherently personal, has different connotations for different people groups and businesses. Up until now, enforcement of privacy was left to the whims of the organisations themselves. But by making these terminologies unequivocally clear, GDPR sets the stage for an unclouded understanding of privacy policies by the data collectors as well as subjects. Some of these are:

\begin{personal}
Any information relating to an person who can be identified, directly or indirectly, in particular by reference to an identifier such as a name, an identification number, location data, online identifier or to one or more factors specific to the physical, physiological, genetic, mental, economic, cultural or social identity of that person.
\end{personal}

\begin{consent}
Any freely given, specific, informed and unambiguous indication of his or her wishes by which the data subject, either by a statement or by a clear affirmative action, signifies agreement to personal data relating to them being processed.
\end{consent}

\begin{profiling}
Any automated processing of personal data to determine certain criteria about a person, in particular to analyse or predict aspects concerning that natural person's performance at work, economic situation, health, personal  preferences,  interests, reliability, behaviour, location or movements.
\end{profiling}

\section*{Information to be made available while collecting data}

The information that must be provided to the subject when the data is being collected, is strictly defined in the GDPR\cite{Galdies}, and this includes:
\begin{enumerate}[noitemsep]
\item the identity and the contact details of the controller and DPO.
\item the purposes of the processing for which the personal data are intended
the legal basis of the processing.
\item where applicable the legitimate interests pursued by the controller or by a third party.
\item where applicable, the recipients or categories of recipients of the personal data.
\item where applicable, that the controller intends to transfer personal data internationally
the period for which the personal data will be stored, or if this is not possible, the criteria used to determine this period.
\item the existence of the right to access, rectify or erase the personal data.
\item the right to data portability.
\item the right to withdraw consent at any time.
\item the right to lodge a complaint to a supervisory authority.

\end{enumerate}

\section*{Technical aspects of storing Personal Data}

Controllers and processors are required to \enquote{implement appropriate technical and organisational measures} taking into account \enquote{the state of the art and the costs of implementation} and \enquote{the nature, scope, context, and purposes of the processing as well as the risk of varying likelihood and severity for the rights and freedoms of individuals.} \cite{Galdies}
The regulation provides specific suggestions for what kinds of security actions might be considered \enquote{appropriate to the risk,} including:
\begin{itemize}
	\item The pseudonymisation and/or encryption of personal data.
	\item The ability to ensure the ongoing confidentiality, integrity, availability and resilience of systems and services processing personal data.
	\item The ability to restore the availability and access to data in a timely manner in the event of a physical or technical incident.
	\item A process for regularly testing, assessing and evaluating the effectiveness of technical and organisational measures for ensuring the security of the processing.
\end{itemize}
Controllers and processors that adhere to either an approved code of conduct or an approved certification may use these tools to demonstrate compliance.

\section*{Profiling}
Full personalisation and other ad serving techniques for example rely on a degree of selection normally built on profiles of behaviour or purchase \cite{Galdies}. According to the GDPR, explicit consent for this is now required.
\begin{itemize}
	\item Individuals have the right not to be subject to the results of automated decision making, including profiling, which produces legal effects on him/her or otherwise significantly affects them. So, individuals can opt out of profiling.
	\item Automated decision making will be legal where individuals have explicitly consented to it, or if profiling is necessary under a contract between an organisation and an individual, or if profiling is authorised by EU or Member State Law
\end{itemize}

\begin{figure}[t!]
	\centering
	\includegraphics[width=0.85\textwidth]{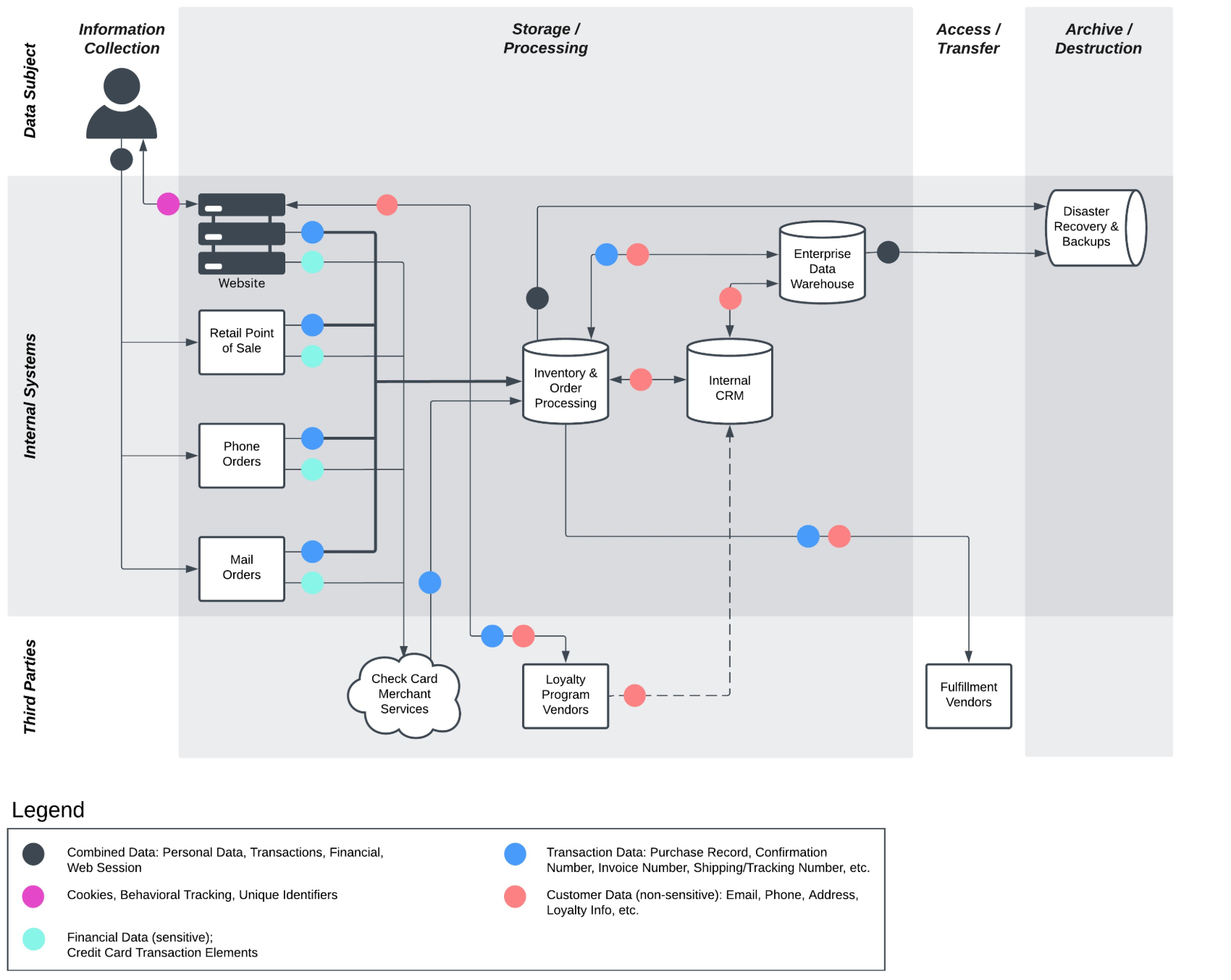}
	\caption{The GDPR Data Flow \cite{Flow}}
	\label{fig:dataflow}
\end{figure}

\section*{Towards GDPR compliance}
One of the first steps for a company towards GDPR compliance, is to a superior understanding of the data flow of the information infrastructure. IT Governance UK \cite{Flowmapping}, in their article, specify the steps that companies should take to ensure this:

\begin{enumerate}
	\item \textbf{Understand the information flow:} An information flow is a transfer of information from one location to another, for example from inside to outside the European Union; or from suppliers and sub-suppliers through to customers.
	\item \textbf{Describe the information flow:} Walk through the information lifecycle to identify unforeseen or unintended uses of data. This also helps to minimise what data is collected. Make sure the people who will be using the information are consulted on the practical implications. Consider the potential future uses of the information collected, even if it is not immediately necessary.
	\item \textbf{Understand the data items:} What kind of data is being processed (name, email, address, etc.) and what category does it fall into (health data, criminal records, location data, etc.)?
	\item \textbf{Formats:} In what format do you store data (hardcopy, digital, database, bring your own device, mobile phones, etc.)?
	\item \textbf{Understand the transfer method:} How do you collect data (post, telephone, social media) and how do you share it internally (within your organisation) and externally (with third parties)?
	\item \textbf{Location, Accountability and Access:} What locations are involved within the data flow (offices, the Cloud, third parties, etc.)? Who is accountable for the personal data? Often this changes as the data moves throughout the organisation? Who has access to the data in question?
\end{enumerate}
\textit{Figure~\ref{fig:dataflow}} provides an example data flow of the organization's infrastructure. Such kind of a modelling becomes essential to understand the governance aspects and the applicability of the GDPR in a fine-grained manner.

\section*{Concluding Remarks}
It remains to be seen how citizens and businesses react to the GDPR. Many businesses have voluntarily extended the coverage of the GDPR to non-European citizens as well, which is certainly a bright spot. The effort itself, after years of deliberations in the EU Parliament, on a subject that is important, and yet so personal, is commendable. To make it successful, we will have to work together and ensure a higher awareness amongst internet-users regarding the GDPR itself, and to use their consent in a judicious and privacy-aware manner. Privacy awareness will surely lead to a safer internet, with lesser instances of spillage of personal information. Our politicians have taken the first step towards providing us some guarantees and control over how our personal information is used. Now, the ball is in our court.

\end{document}